\documentclass[9pt,twocolumn,twoside]{osajnl}
\usepackage[ justification=justified]{caption} 
\usepackage{multirow}
\usepackage{titlesec}
\graphicspath{{images/}{../images/}}
\usepackage{physics}
\journal{optica} 
\setboolean{shortarticle}{false}
\usepackage{lineno}

\title{ High-rate continuous-variable measurement device-independent quantum key distribution with finite-size security}

\author[*]{Adnan A.E. Hajomer}
\author[ ]{Ulrik L Andersen}
\author[$\dagger$]{Tobias Gehring}

\affil[ ]{Center for Macroscopic Quantum States (bigQ), Department of Physics, Technical University of Denmark, 2800 Kongens Lyngby, Denmark}

\affil[ ]{Corresponding authors: * aaeha@dtu.dk, $^\dagger$ tobias.gehring@fysik.dtu.dk}

\begin{abstract}
Building scalable and secure quantum networks requires advanced quantum key distribution (QKD) protocols that support multi-user connectivity. Continuous-variable (CV) measurement-device-independent (MDI) QKD, which eliminates all detector side-channel attacks, is a promising candidate for creating various quantum network topologies—such as quantum access networks and star-type topologies—using standard technology and providing high secure key rates. However, its security has so far only been experimentally demonstrated in asymptotic regimes with limited secret key rates and complex experimental setups, limiting its practical applications. Here, we report the first experimental validation of a CV MDI-QKD system, achieving a secure key rate of 2.6 Mbit/s  against collective attacks in the finite-size regime over a 10 km fiber link. This is achieved using a new system design, incorporating a locally generated local oscillator, a new relay structure, a real-time phase locking system, and a well-designed digital-signal-processing pipeline for quantum state preparation and CV Bell measurements at a symbol rate of 20 MBaud. Our results set a new benchmark for secure key exchange and open the possibility of establishing high-performance CV MDI-QKD networks, paving the way toward a scalable quantum network.

\end{abstract}

\setboolean{displaycopyright}{false}

\begin{document}

\maketitle

\section{Introduction}
\begin{figure*}[h]
\centering
\includegraphics[width=0.8\linewidth]{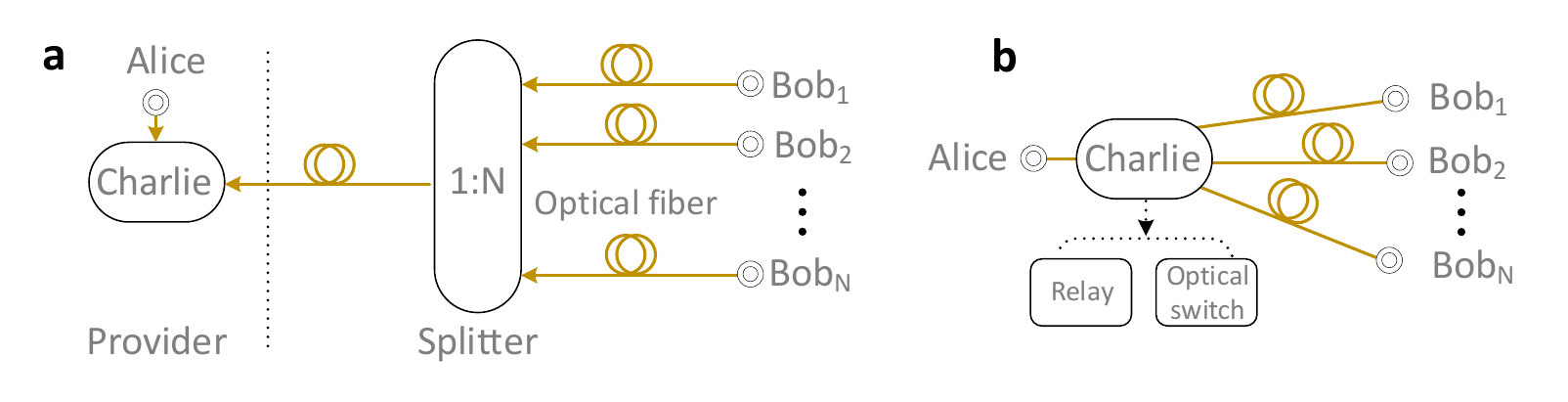}
 \caption{\textbf{Network topologies using asymmetric CV MDI-QKD protocol}. (a)  Upstream quantum access, where Alice and the relay are colocated at the provider station. End users are connected through a quantum channel consisting of a passive beam splitter and optical fiber. (b)  a star-type MDI-QKD network providing users with access to the untrusted party, Charlie, using active optical switch.} 
\label{fig:QAN}
\end{figure*}
 
The distribution of cryptographic keys between remote users is the central problem of secure communication over public channels. Currently, this crucial task is achieved using public-key cryptography, which provides security against attackers with limited computing power~\cite{hellman1976new,rivest1978method}. In contrast, quantum key distribution (QKD) aims to offer information-theoretic secure key exchange without such a restriction of the attackers~\cite{BEN84}. Despite this, in practice, all QKD implementations are vulnerable to side-channel attacks, as real devices behave differently from the theoretical model assumed in the security proof~\cite{xu2020secure,scarani2009security,lydersen2010hacking,gerhardt2011full}.

One solution to prevent side-channel attacks is device-independent (DI) QKD, where the security proof is independent of the physical characteristics of the apparatus~\cite{acin2007device}. This approach has, however, a limited key rate over a reasonable distance~\cite{gisin2010proposal,curty2011heralded,zhang2022device}. As a practical solution, measurement-device-independent (MDI) QKD has been introduced~\cite{braunstein2012side,lo2012measurement,curty2014finite}. This protocol removes the security threat of all known and yet-to-be-discovered side-channel attacks on the detector, which, typically, is the most vulnerable component of a QKD system~\cite{jain2011device,lydersen2010hacking,gerhardt2011full,weier2011quantum,huang2013quantum}. In MDI-QKD, the sources of quantum states are located at the stations of the two trusted communicating parties (Alice and Bob), while correlations are established by the measurement of an untrusted third party (Charlie) located in between Alice and Bob. This allows Alice and Bob to treat \textcolor{black}{the measurement device as a black box, i.e., an untrusted device, }potentially controlled by an eavesdropper (Eve). 

The feasibility of MDI-QKD has been shown through numerous lab demonstrations and field trials, with most implementations using discrete variable or qubit protocols~\cite{yin2016measurement,tang2014field,tang2016measurement,rubenok2013real,wang2022experimental,berrevoets2022deployed}. On the other hand, MDI-QKD protocols, in which the quantum information is encoded in continuous variables (CV), such as the amplitude and phase quadratures of the quantized light field, holds promise for high rates and compatibility with existing telecom infrastructure as they can be realized using standard telecom technologies~\cite{pirandola2015high,li2014continuous}. Despite significant advances in the security proofs of  CV MDI-QKD~\cite{pirandola2015high, li2014continuous,ma2014gaussian, papanastasiou2017finite,zhang2017finite,lupo2018continuous, chen2018composable,ma2019long,wilkinson2020long}, only three proof-of-principle experiments have been conducted to date~\cite{pirandola2015high,tian2022experimental,hajomer2022high}. 

These experiments implemented the asymmetric configuration of the protocol, where the relay was positioned closer to one of the communicating parties. This approach effectively extends the communication distance between Alice and Bob by mitigating the inherent trade-off between the relay's location and the maximum achievable distance~\cite{pirandola2015high}. Additionally, the asymmetric configuration can be used to implement various quantum network topologies~\cite{frohlich2015quantum, wang2023experimental}. Figure~\ref{fig:QAN} (a) illustrates how the asymmetric CV MDI-QKD protocol can be applied to an upstream quantum access network~\cite{frohlich2015quantum}, where the provider station (with Alice and Charlie colocated) connects to end users (Bobs) via a passive beam splitter and fiber channels. Each Bob establishes a key with the provider through either time or frequency multiplexing. Furthermore, a star-type quantum network~\cite{wang2023experimental} can also be realized using this asymmetric structure along with optical switching, as shown in Fig.~\ref{fig:QAN} (b). For more details on the network topologies of asymmetric CV MDI-QKD, we refer the reader to the supplementary information in Ref.~\cite{pirandola2015high}.

The previous experiments mentioned above relied heavily on traditional technologies, resulting in complex setups that limited both the system's speed and its practical applications, hindering the exploration of the technology in complex quantum network designs. For instance, they required transmitting the local oscillator through the same fiber and employed minimal digital signal processing (DSP). Furthermore, none of these demonstrations considered finite-size effects, an essential aspect for real-world applications.

\begin{figure*}[h]
\centering
\includegraphics[width=0.8\linewidth]{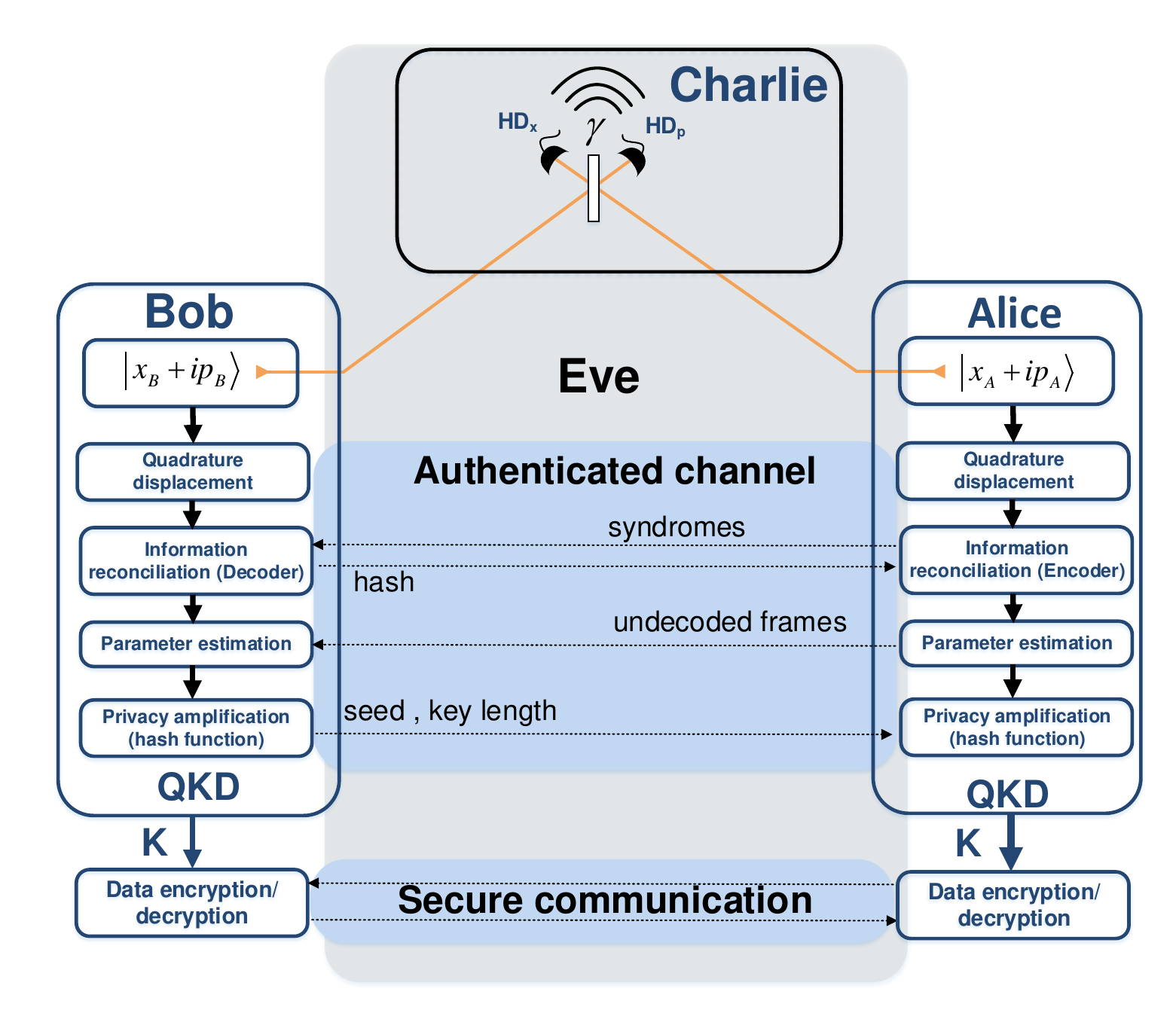}
 \caption{\textbf{CV MDI-QKD protocol}. Alice and Bob prepare coherent states and send them to an untrusted relay, Charlie, who  performs \textcolor{black}{continuous variable Bell state measurements. Based on the outcomes of these measurements, Alice and Bob establish quadrature amplitude correlations for keys generation}. The key distribution protocol involves multiple phases, including information reconciliation, parameter estimation, and privacy amplification. These  steps are executed  over an authenticated channel. \textcolor{black}{Upon completing the protocol, Alice and Bob acquire symmetric keys, denoted as $K$, which can be used for  data encryption and decryption}.} 
\label{fig:1}
\end{figure*}

Here, we report on an experimental demonstration of CV MDI-QKD achieving a secure key rate of 2.6 Mbit/s against collective attacks while accounting for finite-size effects. This was enabled by a new system design, which simplifies the relay structure for the CV Bell state measurement (BSM), using a polarization-based 90-degree hybrid and a low-cost field-programmable gate array (FPGA) for real-time phase locking~\cite{suleiman202240}. Unlike previous experiments ~\cite{tian2022experimental,hajomer2022high}, our setup utilized the local local oscillator (LLO) scheme by leveraging the asymmetric configuration of the protocol,  along with wavelength and polarization multiplexing techniques. This eliminates the need for an extra fiber channel for phase locking.  Furthermore, we have replaced conventional technologies such as pulse carving, time-multiplexing, and analog propagation delay compensation with efficient digital signal processing (DSP) pipelines, enabling a reconfigurable system and simplified optical subsystem. This advancement allowed our system to operate at a symbol rate of 20 MBaud, representing a more than one order improvement over previous experiments~\cite{tian2022experimental,pirandola2015high}. Finally, we demonstrated the coexistence of classical and quantum signals in the same fiber channel. Introducing the DSP approach to MDI-QKD is the first milestone towards implementing CV-QKD in complex quantum networks, where minimal trust in the devices is required.

 \begin{figure*}[h]
\centering
\includegraphics[width=\linewidth]{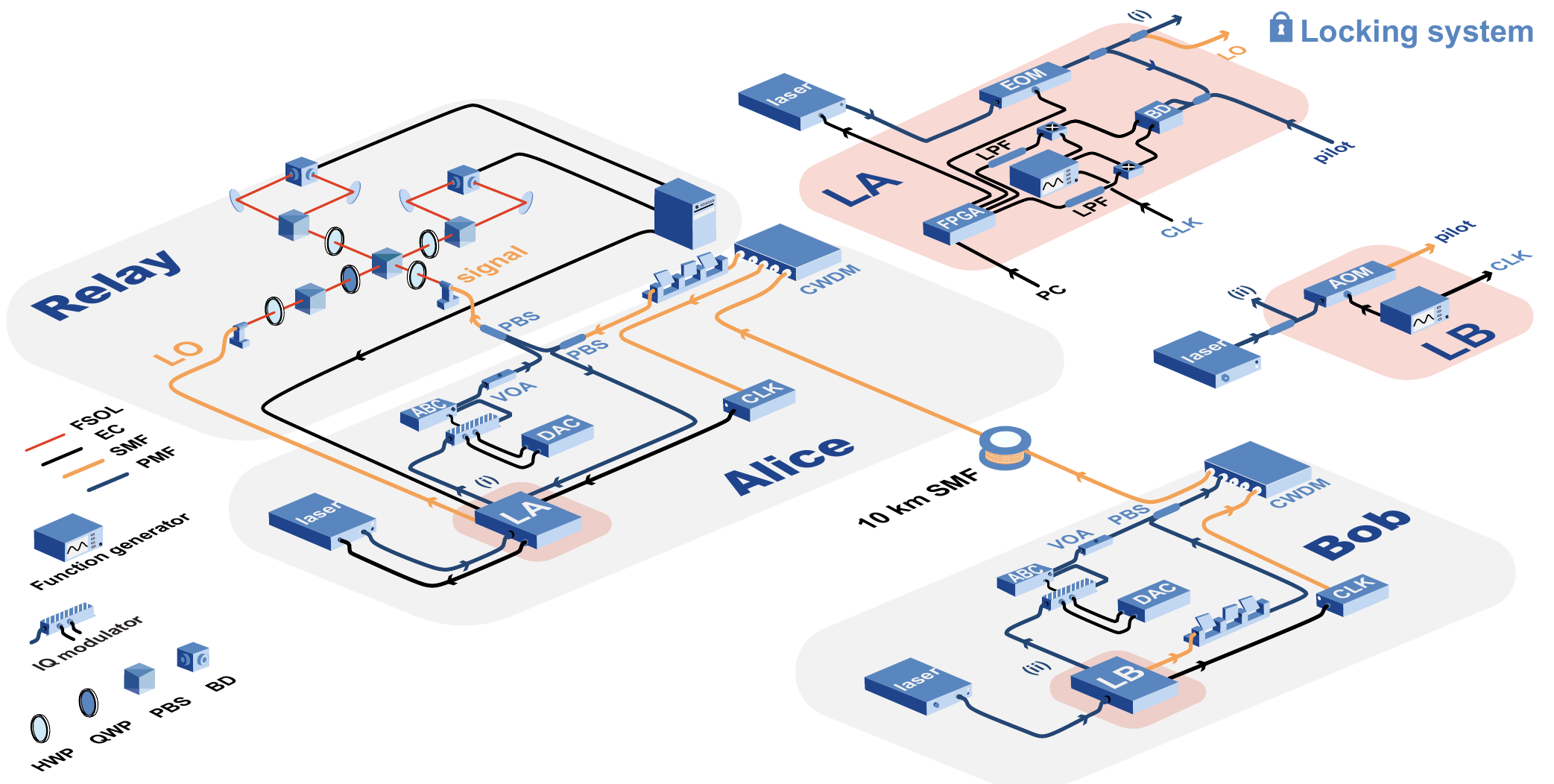}
\caption{\textbf{CV MDI-QKD system}. The relay is co-located with Alice's station. The sender stations were constructed using polarization-maintaining fiber (PMF) components (purple fiber), while the relay utilizes free-space optical components. Alice and Bob prepare an ensemble of Gaussian-modulated coherent states by modulating phase-locked 1550 nm continuous-wave laser beams using an IQ modulator driven by a digital-to-analog converter (DAC) and an automatic bias controller (ABC). A variable optical attenuator (VOA) was used after each IQ modulator to adjust the modulation variance of the quantum signal. To synchronize the clocks of Alice's and Bob's stations, an optical clock (CLK) at 1310 nm was generated at Bob's station and sent to Alice after being wavelength-multiplexed with the quantum signal using Coarse Wavelength Division Multiplexing (CWDM). The two lasers were phase-locked using a heterodyne phase-locking technique involving two synchronized function generators . One generator is located at Bob's station and drives an acousto-optic modulator (AOM), while the other is located at Alice's station and used as a reference local oscillator for phase detection. Finally, a field-programmable gate array (FPGA) was used to drive the piezoelectric wavelength modulator of the laser and the electro-optical modulator (EOM) for frequency and phase control, respectively. FSOL: free-space optical line; EC: electronic connection; LPF: \textcolor{black}{low-pass filter; LA: Locking system components at Alice's site;  LB: Locking system components   at Bob's site.}} 
\label{fig:2}
\end{figure*}

\section{CV MDI-QKD Protocol}
Figure~\ref{fig:1} illustrates the full protocol of secure communication enabled by CV MDI-QKD, which consists of three phases: the quantum phase, data processing, and data encryption. 

During the quantum phase, Alice and Bob independently prepare coherent states, denoted as $\ket{x_A+ip_A}$ and $\ket{x_B+ip_B}$, respectively, whose amplitudes are drawn from independent and identical zero-mean Gaussian distributions. The real (imaginary) part of the
complex number is equal to the amplitude (phase) quadrature component in the phase space, with quadrature operators fulfilling the commutation relation $[\hat{x},\hat{p}]=2i$. These states are then transmitted through separate quantum channels to a detection station (Charlie), which is under the full control of Eve. Assuming a faithful operation of the relay, Charlie performs CV BSM by overlapping the incoming states in a balanced beam splitter, followed by double homodyne detection (HD), which measures the conjugate quadrature operator pair $\hat{x}$ and $\hat{p}$. Finally,  Charlie publicly announces the measurement outcome, $\gamma = \gamma_x+i\gamma_p$, which enables Alice and Bob to establish posteriori correlations. 

The data processing begins with Bob using the outcome of the measurement $\gamma$ to infer Alice's variable by performing quadrature displacement, resulting in 
$x_A+\delta_x=\gamma_x-\rho x_B$  and  $p_A+\delta_p=\gamma_p+\zeta p_B$, where $\delta_x$ and $\delta_p$ account for the detection noise and $\rho$ and $\zeta$ are scaling factors, that can be computed as described in~\cite{lupo2018continuous}. Alice and Bob then perform information reconciliation by exchanging information, such as error correction syndromes and hashing values for the correctness test, over an authenticated channel that Eve can monitor but not alter. This is followed by parameter estimation to evaluate the information advantage of the communicating parties over Eve and privacy amplification to generate the final secret key. Unlike the conventional protocol, performing parameter estimation after information reconciliation allows the communicating users to use all the measurements for parameter estimation and key generation~\cite{lupo2018parameter, jain2022practical}. After going through these steps, Alice and Bob share secret keys, $K$, which can be used for a cryptographic application, for instance data encryption and decryption.

\section{CV MDI-QKD system}
 The schematic of our CV MDI-QKD system is shown in  Fig.~\ref{fig:2}. The system consisted of the two sender stations (Alice and Bob) and the relay (Charlie).  We implemented the asymmetric configuration of the MDI protocol, where Alice and the relay are colocated nearby, a setup that extends the system's communication range \cite{pirandola2015high}. Bob's transmitter and the relay were connected via a quantum channel made of single-mode fiber (SMF). 
 The following subsections will thoroughly explain each station's optical layout and associated digital signal processing (DSP). 
 
\subsection{Senders (Alice and Bob)}
Alice's and Bob's stations each have a continuous-wave (CW) laser operating at 1550 nm with a linewidth of $\approx$ 100 Hz. These independent lasers were phase-locked using a digital locking system~\cite{suleiman202240}. In each station, \textcolor{black}{baseband modulation} was performed using an in-phase and quadrature (IQ) modulator driven by a dual-channel digital-to-analog converter (DAC) with 16-bit resolution and a sampling rate of 1 GSample/s~\cite{hajomer2022modulation}. The IQ modulators were operated in carrier suppression mode, with the  bias voltages controlled by automatic bias controllers (ABCs). After the IQ modulators, variable optical attenuators (VOAs) were used to adjust the variance of the generated thermal states. Alice's source and the relay were co-located and connected to Bob's transmitter with a 10 km long SMF. 

\begin{figure}[t]
\centering
\includegraphics[width=\linewidth]{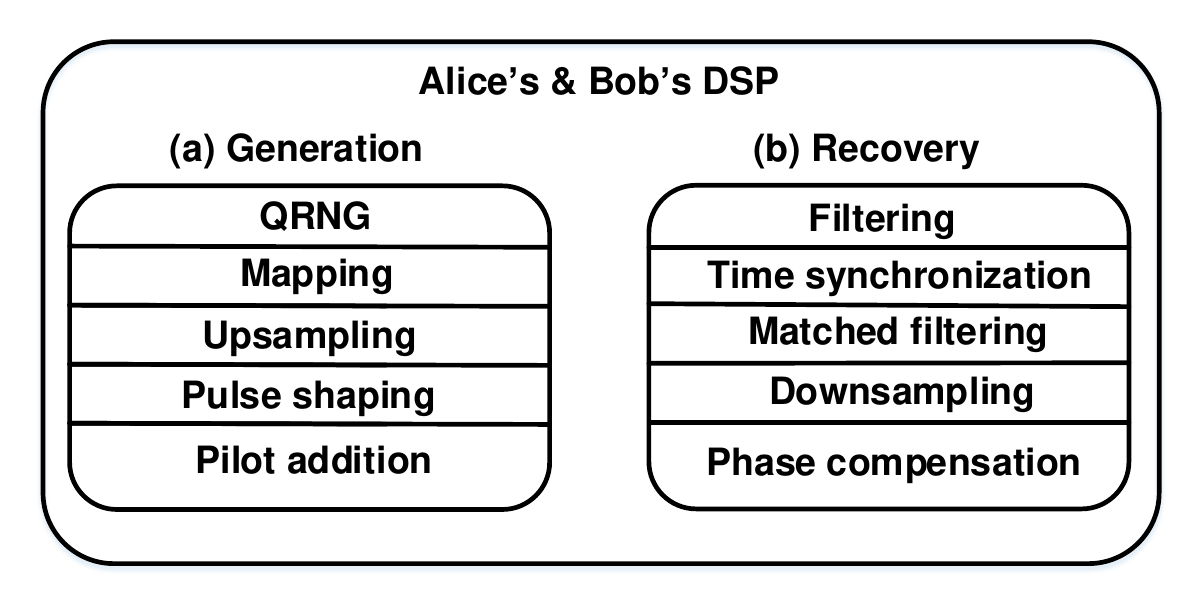}
\caption{\textbf{Digital signal processing (DSP) chain}. DSP pipeline for (a) quantum state preparation; (b) quantum symbol recovery. See the main text for the details.} 
\label{fig:3}
\end{figure}

Alice and Bob digitally generated driving waveforms using the DSP routine in Fig.~\ref{fig:3} (a) to produce an ensemble of coherent states. The complex amplitude of the coherent states was formed by drawing random numbers from independent and identical zero-mean Gaussian distributions, obtained by transforming the uniform distribution of a vacuum fluctuation-based quantum random number generator (QRNG)~\cite{gehring2021homodyne}. These quantum symbols were drawn at a symbol rate of 20 MBaud and up-sampled to 1 GSample/s after which they were pulse-shaped using a root-raised cosine (RRC) filter with a roll-off of 0.2. This eliminated the need for additional amplitude modulation for pulse carving ~\cite{tian2022experimental}. To this baseband signal, Bob multiplexed in frequency a pilot tone at 15 MHz to estimate the residual phase noise of the locking system. Alice and Bob then uploaded their driving waveforms to their DACs to obtain the corresponding electrical signals that drove the IQ modulators.

 Synchronizing clocks between two remote stations is crucial for practical CV MDI-QKD systems. In our system, Alice's and Bob's stations were clock synchronized using the following techniques. Initially, a 10 MHz master clock was generated at Bob's station. This clock was then converted to an optical signal with a power of 1 mW at a wavelength of 1310 nm using an electrical-to-optical converter circuit.  Subsequently, we multiplexed the optical clock with the quantum signal using a Coarse Wavelength Division Multiplexer (CWDM) and transmitted it to Alice's station through the same fiber channel. This approach significantly simplifies the system's structure compared to previous experiments~\cite{tian2022experimental,hajomer2022high} and demonstrates the feasibility of quantum and telecommunication channels coexisting on the same infrastructure. At Alice's station. The optical clock was then converted back to an electrical signal using an optical-to-electrical converter circuit. This signal was then distributed to the DAC, the locking system and the relay's analog-to-digital converters (ADCs).  As a result, digital clock synchronization~\cite{chin2022digital} between the transmitters and the relay was not necessary for CV BSM.
 
\subsection{Phase locking}
 To experimentally realize CV BSM over a long distance, it is crucial to phase-lock the two separate lasers used to generate the quantum states at Alice's and Bob's stations.This ensures interference between Alice's and Bob's generated quantum states. Precise control over the relative phase between the quantum states and the LO is essential for enabling dual-homodyne detection. We implemented these features using a heterodyne optical locking system, a technique that is well-established~\cite{minder2019experimental,bordonalli1999high,suleiman202240}. 

\textcolor{black}{The inset of Fig. 2 illustrates the components of the locking system located  at Alice's station (LA) and Bob's station (LB), including the two independent lasers employed for quantum state preparation. In this setup, Alice's laser (slave) was synchronized with Bob's laser (master). This arrangement allowed the locking loop to  compensate for fast phase fluctuations caused by the long fiber link between Bob's station and the relay.} 

\textcolor{black}{At Bob's station, a portion of the laser beam was frequency-shifted by 40 MHz using an acousto-optical modulator (AOM) driven by a function generator.  The  frequency-shifted beam was then sent to Alice's  station, where it was interfered with small part of Alice's laser beam on a balanced beam splitter. This interference produced  a beat signal, which was detected by a balanced detector. The phase detection was then performed by analog I-Q demodulation at 40 MHz. An FPGA was used to generate an error signal and to implement a proportional-integral (PI) controller. The PI controller drove a piezoelectric wavelength modulator inside Alice's laser to compensate for slow phase fluctuations and an electro-optic phase modulator (EOM) to compensate for fast ones. Finally, Alice's stabilized laser was used as an optical source and shared with the relay as LO through a short fiber channel. Notably, within the framework of the MDI-QKD protocol, which assumes that Eve controls the measurement device, the fiber channel used for distributing the laser beam between Alice and the relay could also, in principle, be susceptible to Eve's control, despite being physically located inside Alice's trusted station.} More details about the locking system can be found in~\cite{suleiman202240}.

\subsection{Relay (Charlie)}
 For CV BSM, the relay station typically consists of a balanced beamsplitter for overlapping Alice's with Bob's quantum signal, and employs double homodyne detection for measuring the orthogonal quadratures~\cite{pirandola2015high,tian2022experimental}. This setup necessitates the generation of two LOs with a phase difference of $\pi/2$, typically achieved using an additional phase-locking system to precisely control the phase difference between the two LO arms. However, this adds a layer of complexity to the system's structure. 
 
 To simplify the system design, we developed a new relay structure that simultaneously measures the conjugate quadratures without phase-locking~\cite{hajomer2022high}, as shown in Fig.~\ref{fig:2}. This relay leverages  the principle of polarization-based 90-degree hybrids, in which the phase shift of $\pi/2$ is introduced by preparing the LO in circular polarization using a quarter wave-plate (QWP) and splitting it into two arms using a polarization beamsplitter (PBS)~\cite{kazovsky1989phase}. \textcolor{black}{For further details about our detection scheme see supplementary materials}. 
 
 In our  configuration, the incoming beams from Alice and Bob were overlapped at a fiber-based PBS and then free-space coupled into the hybrid. \textcolor{black}{As the performance of CV MDI-QKD relies on relay efficiency, the hybrid was built from free-space bulk components and balanced detectors equipped with photodiodes that have a high quantum efficiency of  $99\%$. These detectors offer a bandwidth of approximately 50 MHz and ensure linear operation up to around 13 MHz}.  After accounting for the visibility of the interference and the insertion loss, the total quantum efficiency of the relay was measured to be $94\%$. The detectors' output was digitized using a dual-channel ADC with a sampling rate of 1 GSample/s, which was clock synchronized with Alice’s and Bob's DACs. 

\subsection{Data processing}
In our experiment, we carried out two types of measurements: one-time shot noise calibration~\cite{zhang2020one}, in which the signal path was blocked while keeping the LO open, and CV BSM. \textcolor{black}{  These measurements were carried out consecutively and autonomously, without the need for user intervention, using a software framework written in Python. The measurement time was divided into frames, each with $10^7$ samples.} To determine the modulation variance of the thermal states at the output of each transmitter, we conducted back-to-back measurements, where one station at a time was connected to the relay through a short fiber channel, and the VOA was finely tuned to set the modulation variance of the thermal states. 

To create correlated random variables for secret key generation, Alice and Bob first applied the DSP pipeline shown in Fig.~\ref{fig:3} (b). \textcolor{black}{The first step involved} low-pass filtering the relay output to eliminate the pilot tone at 15 MHz. Subsequently, they accounted for the propagation delays through fiber channels and various electronic components by calculating the cross-correlation between the transmitted samples and the relay output. Unlike one-way CVQKD, temporal synchronization was done on the transmitted samples instead of the relay output since Alice and Bob were connected to the relay through fiber channels with different lengths. \textcolor{black}{Following this}, the same Root Raised Cosine (RRC) filter used in waveform generation was applied as a matched filter to downsample the synchronized relay output samples ($\gamma$). As Alice's laser was shared with the relay as LO through a short fiber channel, Alice and Bob encountered slow phase fluctuations \textcolor{black}{relative to} the LO. To compensate for \textcolor{black}{these fluctuations}, Alice and Bob rotated their symbols using the relay output as reference symbols to maximize the correlation as, 
\begin{align}
\hat{\alpha}_\text{Alice} =  \underset{\theta_1}{\arg\max}~Cov(\alpha_\text{Alice}\exp(j\theta_1),\gamma) \nonumber\\
\hat{\alpha}_\text{Bob} =  \underset{\theta_2}{\arg\max}~Cov(\alpha_\text{Bob}\exp(j\theta_2),\gamma),
\label{Eq:1}
\end{align} 
where $\hat{\alpha}$ is the rotated symbol, $\theta_x$ is the bulk phase and $Cov(\cdot)$ denotes the covariance. \textcolor{black}{The  filtering process was the most time-consuming step within this DSP chain}. After the DSP, Alice and Bob performed quadrature displacement. 


 \begin{figure}[t]
\centering
\includegraphics[width=\linewidth]{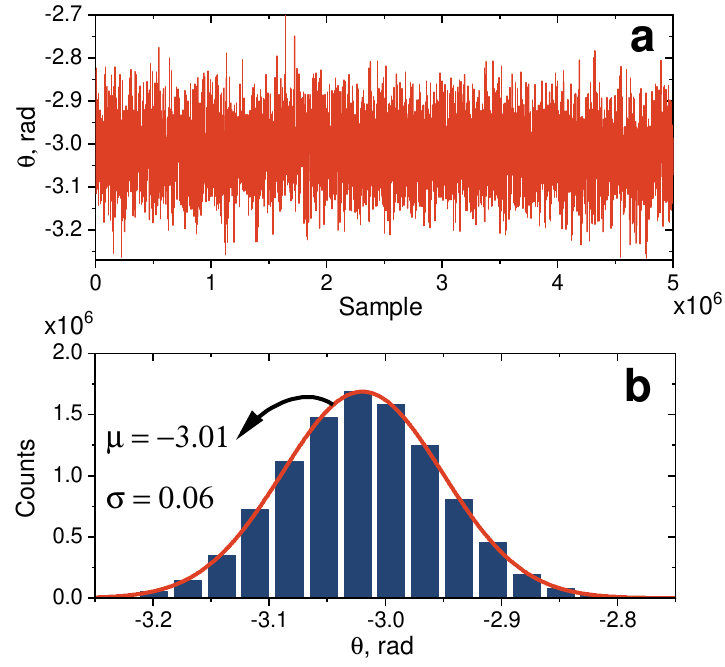}
\caption{\textbf{Phase noise measurement}. (a) The unwrapped phase of pilot tone after phase locking. (b) The histogram of stabilized phase drift with mean $\mu$ close to -3 rad and standard deviation $\sigma$ of 0.06 rad.} 
\label{fig:4}
\end{figure}

\begin{table*}[t]
\centering
\caption{\textbf{ Experimental parameters for secret key \textcolor{black}{rate} calculation}. $V_A$ and $V_B$: Alice and Bob modulation variance, B: symbol rate, N: total number of symbols, $\xi$: excess noise, $\tau_A$ and $\tau_B$: Alice and Bob transmissivities, $\eta$: the relay efficiency, $\beta$: IR efficiency.} 
\resizebox{0.8\hsize}{!}{
\begin{tabular}{ccccccccc}
\hline
$V_A$& $V_B$& B& N& $\xi$& $\tau_A$& $\tau_B$& $\eta$& $\beta$ \\
\hline
6.5 SNU&  6.5 SNU& 20 Mbaud& $4\times10^6$& 39.5 mSNU& 1&  0.56& 0.94 &97\%\\
\hline
\end{tabular}
}
  \label{tab:1}
\end{table*}

\section{Results}
\textcolor{black}{The initial} step in implementing the CV MDI-QKD system involves \textcolor{black}{equipping} users with phase-locked laser beams. This ensures interference and measurement of conjugate quadratures at the relay. Assessing the locking system's efficacy is crucial for evaluating the overall system performance. \textcolor{black}{We achieved} this by measuring the residual phase noise associated with the locking system, using a frequency multiplexed pilot tone with Bob's quantum signal. \textcolor{black}{Figure~\ref{fig:4} a(b)} depicts the unwrapped phase of the pilot tone and its histogram, respectively. The Gaussian distribution of the unwrapped phase confirms the locking system's success in compensating for phase noise, modeled as a random-walk process~\cite{chin2021machine}. The average phase noise value of $\approx$ -3 rad  indicates a slow phase shift between the LO and Alice's and Bob's signals, likely due to a portion of Alice's locked laser being shared as LO with the relay through a short fiber \textcolor{black}{channel}. This residual phase, which slowly changes in time, was compensated for during DSP. The residual phase noise, defined as the standard deviation of the unwrapped phase, was 0.06 rad, equivalent to a phase error of $3.44^\circ$. This value seems to be within the typical range of phase error for optical phase-locking systems, compared to \textcolor{black}{prior} experiments~\cite{suleiman202240,minder2019experimental}.    

 Residual phase noise primarily contributes to so-called excess noise, which limits the rate and distance of CV MDI-QKD.  In our Gaussian-modulated coherent state-based system, the excess noise due to the residual phase noise can be calculated as~\cite{marie2017self} \textcolor{black}{$\xi_{\text{phase}}=2TV_{\text{}}\left(1-e^{-\frac{{\sigma_\theta}^2}{2}}\right)$, where $T$ is the transmittance, including the quantum channel and the relay efficiency};  ${\sigma_\theta}^2$ is the variance of the residual phase noise and $V_\text{}$ is the modulation variance. \textcolor{black}{Enhancing system performance involves minimizing, $\xi_{\text{phase}}$, achievable by reducing} the phase error via an improved locking system or operating at a lower modulation variance. Therefore, we \textcolor{black}{operated} our system at a low modulation variance of 6.5 shot-noise units (SNUs), resulting in $\xi_{\text{phase}} \approx$ \textcolor{black}{12.6 mSNU}, considering $\sigma_\theta = 0.06$ rad.

\begin{figure}[t]
\centering
\includegraphics[width=\linewidth]{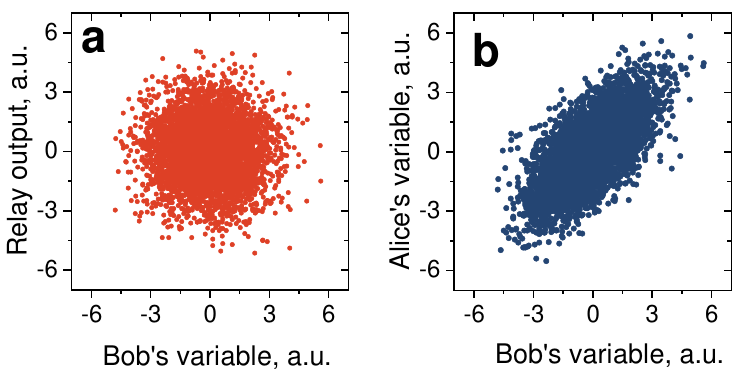}
\caption{\textbf{\textcolor{black}{Analysis of} posteriori correlations}. The correlations  between the relay output (a), Alice's symbols (b), and Bob's symbols following quadrature displacement.} 
\label{fig:5}
\end{figure}

\textcolor{black}{Our} proposed relay structure is one of the basic building blocks of our CV MDI-QKD implementation. To assess its effectiveness, we evaluated  the information advantage Bob gains  in attempting to infer Alice's variable using the relay output. Figure~\ref{fig:5}~a(b) illustrates the correlations between relay outcome (Alice's symbols) and Bob's symbols after quadrature displacement. \textcolor{black}{Notably, the displacement operation enhances} the correlation between Alice's and Bob's symbols compared to the correlation between Bob's symbols and the relay outcome. This enhancement enables Alice and Bob to establish posteriori correlations through the relay output, demonstrating the relay's capability in facilitating CV BSM.

\textcolor{black}{Finally, to validate the feasibility of our CV MDI-QKD system, we demonstrated positive secret key rates over a 10 km fiber channel between Alice/relay and Bob, both in the asymptotic and finite-size regimes}. The secret key rate and parameter estimation were computed based on the methods outlined in~\cite{pirandola2015high,papanastasiou2017finite}. For further details about secret key rate calculation, see the supplementary materials. A summary of the main parameters used for \textcolor{black}{secret key rate evaluation} can be found in Table \ref{tab:1}. Alice and Bob generated an ensemble of $4\times10^6$ coherent states at a symbol rate of 20 Mbaud,  with a modulation variance of 6.5 SNU, transmitting them through channels with transmissivities of 1 and 0.56, respectively. During parameter estimation, Bob used all symbols, \textcolor{black}{assuming that information reconciliation is performed prior to parameter estimation}. After CV BSM, an excess noise of 39.5 mSNU was measured at the relay. For a fair comparison with the previous works~\cite{pirandola2015high,tian2022experimental}, we assumed an information reconciliation efficiency $\beta$ of 97\%.  

\begin{figure}[t]
\centering
\includegraphics[width=\linewidth]{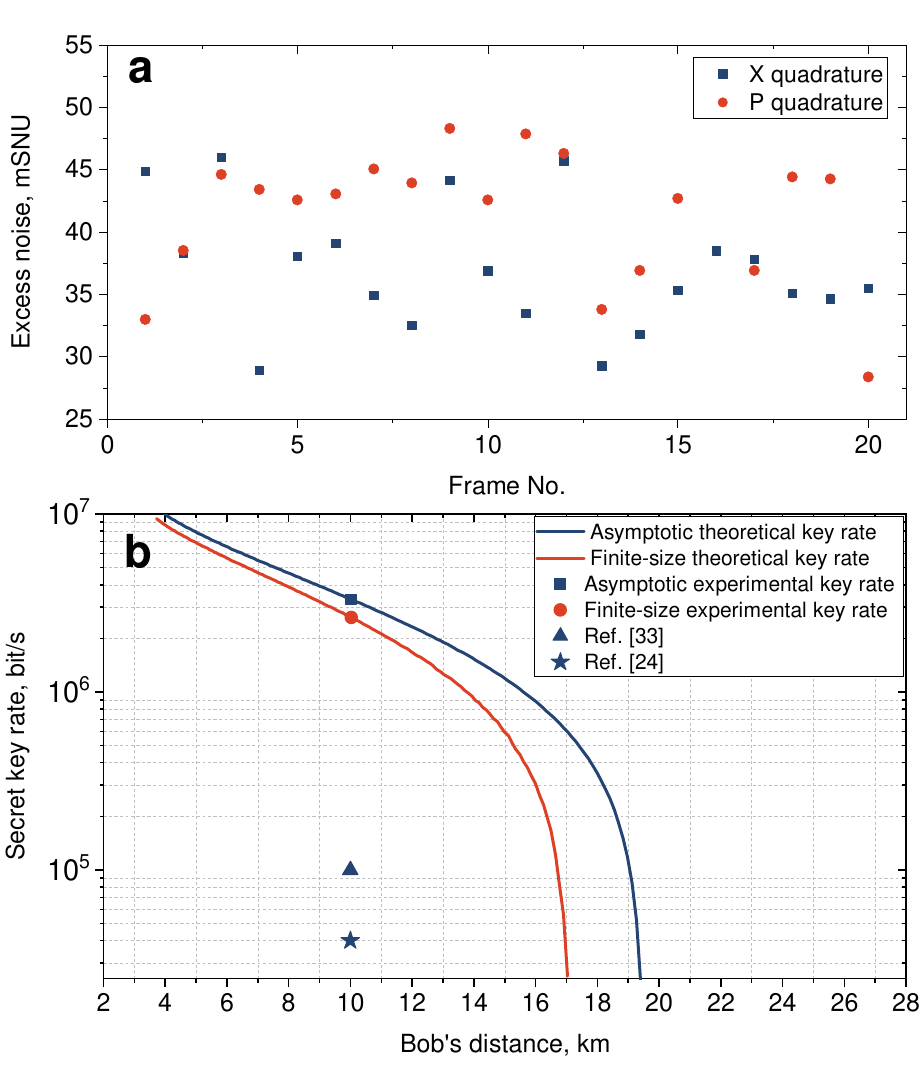}
\caption{\textbf{Performance of our CV MDI-QKD system}. (a) Measured excess noise. (b) \textcolor{black}{Expected secret key rate}.  } 
\label{fig:6}
\end{figure}

Figure~\ref{fig:6} (a) depicts the measured excess noise at the relay for different frames. The average excess noise was $\approx$ 39.5 mSNU, while the worst-case estimator consider finite-size correction was $\approx$ 45 mSNU.This estimate was computed using a Gaussian confidence interval~\cite{papanastasiou2017finite} \textcolor{black}{(see the supplementary material for further details)}. As one-time shot noise calibration was used, the electronic noise of the detector is not subtracted from excess noise. Similarly, Figure~\ref{fig:6} (b) compares the experimental secret key rates in asymptotic and finite-size regimes with numerical simulations based on experimental parameters. With a block size of $4\times10^6$ and a failure probability of $10^{-10}$, we achieved a positive \textcolor{black}{expected} secrete key rate of 2.6 Mbit/s. 

\section{Discussion}
MDI-QKD is an effective approach to eliminate all detector side-channel vulnerabilities. In particular, CV MDI-QKD benefits from its compatibility with standard telecom components and has the potential for higher key rates over metropolitan distances compared to qubit-based protocols~\cite{pirandola2015high}. However, experimental validation of high-rate CV MDI-QKD with finite-key security had yet to be demonstrated, which is essential for practical applications. In this work, we reported the first experimental demonstration of CV MDI-QKD achieving a  key rate of 2.6 Mbit/s secure against collective attacks, while accounting for finite-size effects over a 10 km fiber channel. This achievement was made possible through the development of a DSP-based CV MDI-QKD system operating at a symbol rate of 20 Mbaud.

Our implementation demonstrates a more than tenfold improvement in key rate at the same distance and information reconciliation efficiency compared to previous CV MDI-QKD experiments~\cite{pirandola2015high,tian2022experimental}, which only reported asymptotic key rates, as shown in Fig.~\ref{fig:6}. Additionally, we eliminated the need for traditional techniques such as pulse carving, transmitted local oscillators, and time multiplexing. Instead, our system used reconfigurable DSP for quantum state preparation and CV BSM.

In our system implementation we located the relay of CV MDI-QKD at Alice's station. This introduces some resemblance to one-sided device-independent (1sDI) QKD systems. However, the fundamental difference between the two systems in terms of the measurement results being public or private remains. Furthermore, comparing the performance of CV-MDI-QKD and CV 1sDI-QKD, reveals a superior performance for CV MDI-QKD~\cite{walk2016experimental, gehring2015implementation} which is due to the use of the entropic uncertainty relation in 1sDI security proofs.

While our current implementation offers the highest key rate to date and a simple system design, further improvements are possible to achieve composable security and extend operational distance. To meet composable security requirements, a higher symbol rate is necessary, which could be achieved by increasing the detector bandwidth~\cite{bruynsteen2021integrated,hajomer2024continuous}. Reducing losses and excess noise would also extend the system’s range, potentially through photonic chip integration for the transmitter and relay, as well as employing machine learning techniques for phase noise compensation~\cite{chin2021machine}. Machine learning could further enhance our DSP system by enabling the use of lower-cost lasers with broader linewidths. Additionally, physically separating the relay could benefit certain network configurations, though it will come at the expense of a reduced secure distance~\cite{pirandola2015high}. Our design can accommodate this by incorporating intradyne detection with an independent laser, which could also be used to lock Alice and Bob’s lasers for Bell measurement. For practical secret key generation, information reconciliation and privacy amplification must be integrated.

In conclusion, this experiment has the potential to enable the development of key quantum network topologies, paving the way for scalable and secure quantum networks.    

\begin{backmatter}

\bmsection{Acknowledgments} We thank Iyad Suleiman for the discussions regarding the phase locking system, Panagiotis Papanastasiou and Stefano Pirandola for the discussions about security analysis and quadrature displacement, \textcolor{black}{and Ivan Derkach for designing the schematic of the system}. This work was supported by the European Union’s Horizon 2020 research and innovation programs through CiViQ (grant agreement no. 820466) and OPENQKD (grant agreement no.\ 857156). We also acknowledge support from Innovation Fund Denmark (CryptQ, 0175-00018A) and the Danish National Research Foundation, Center for Macroscopic Quantum States (bigQ, DNRF142).

\smallskip

\bmsection{Data Availability Statement} 
Data underlying the results presented in this paper are available from the authors upon reasonable request.

\smallskip
\bmsection{Competing interests} The authors declare no competing interests.

\end{backmatter}

\bibliography{sample}

\begin{thebibliography}{10}
\newcommand{\enquote}[1]{``#1''}

\bibitem{hellman1976new}
M.~Hellman, \enquote{New directions in cryptography,} {\protect\JournalTitle{IEEE transactions on Information Theory}} \textbf{22}, 644--654 (1976).

\bibitem{rivest1978method}
R.~L. Rivest, A.~Shamir, and L.~Adleman, \enquote{A method for obtaining digital signatures and public-key cryptosystems,} {\protect\JournalTitle{Communications of the ACM}} \textbf{21}, 120--126 (1978).

\bibitem{BEN84}
C.~H. Bennett and G.~Brassard, \enquote{{Quantum cryptography: Public key distribution and coin tossing},} in \emph{Proceedings of IEEE International Conference on Computers, Systems, and Signal Processing,}  (India, 1984), p. 175.

\bibitem{xu2020secure}
F.~Xu, X.~Ma, Q.~Zhang, H.-K. Lo, and J.-W. Pan, \enquote{Secure quantum key distribution with realistic devices,} {\protect\JournalTitle{Reviews of Modern Physics}} \textbf{92}, 025002 (2020).

\bibitem{scarani2009security}
V.~Scarani, H.~Bechmann-Pasquinucci, N.~J. Cerf, M.~Du{\v{s}}ek, N.~L{\"u}tkenhaus, and M.~Peev, \enquote{The security of practical quantum key distribution,} {\protect\JournalTitle{Reviews of modern physics}} \textbf{81}, 1301 (2009).

\bibitem{lydersen2010hacking}
L.~Lydersen, C.~Wiechers, C.~Wittmann, D.~Elser, J.~Skaar, and V.~Makarov, \enquote{Hacking commercial quantum cryptography systems by tailored bright illumination,} {\protect\JournalTitle{Nature photonics}} \textbf{4}, 686--689 (2010).

\bibitem{gerhardt2011full}
I.~Gerhardt, Q.~Liu, A.~Lamas-Linares, J.~Skaar, C.~Kurtsiefer, and V.~Makarov, \enquote{Full-field implementation of a perfect eavesdropper on a quantum cryptography system,} {\protect\JournalTitle{Nature communications}} \textbf{2}, 349 (2011).

\bibitem{acin2007device}
A.~Ac{\'\i}n, N.~Brunner, N.~Gisin, S.~Massar, S.~Pironio, and V.~Scarani, \enquote{Device-independent security of quantum cryptography against collective attacks,} {\protect\JournalTitle{Physical Review Letters}} \textbf{98}, 230501 (2007).

\bibitem{gisin2010proposal}
N.~Gisin, S.~Pironio, and N.~Sangouard, \enquote{Proposal for implementing device-independent quantum key distribution based on a heralded qubit amplifier,} {\protect\JournalTitle{Physical review letters}} \textbf{105}, 070501 (2010).

\bibitem{curty2011heralded}
M.~Curty and T.~Moroder, \enquote{Heralded-qubit amplifiers for practical device-independent quantum key distribution,} {\protect\JournalTitle{Physical Review A}} \textbf{84}, 010304 (2011).

\bibitem{zhang2022device}
W.~Zhang, T.~van Leent, K.~Redeker, R.~Garthoff, R.~Schwonnek, F.~Fertig, S.~Eppelt, W.~Rosenfeld, V.~Scarani, C.~C.-W. Lim \emph{et~al.}, \enquote{A device-independent quantum key distribution system for distant users,} {\protect\JournalTitle{Nature}} \textbf{607}, 687--691 (2022).

\bibitem{braunstein2012side}
S.~L. Braunstein and S.~Pirandola, \enquote{Side-channel-free quantum key distribution,} {\protect\JournalTitle{Physical review letters}} \textbf{108}, 130502 (2012).

\bibitem{lo2012measurement}
H.-K. Lo, M.~Curty, and B.~Qi, \enquote{Measurement-device-independent quantum key distribution,} {\protect\JournalTitle{Physical review letters}} \textbf{108}, 130503 (2012).

\bibitem{curty2014finite}
M.~Curty, F.~Xu, W.~Cui, C.~C.~W. Lim, K.~Tamaki, and H.-K. Lo, \enquote{Finite-key analysis for measurement-device-independent quantum key distribution,} {\protect\JournalTitle{Nature communications}} \textbf{5}, 3732 (2014).

\bibitem{jain2011device}
N.~Jain, C.~Wittmann, L.~Lydersen, C.~Wiechers, D.~Elser, C.~Marquardt, V.~Makarov, and G.~Leuchs, \enquote{Device calibration impacts security of quantum key distribution,} {\protect\JournalTitle{Physical Review Letters}} \textbf{107}, 110501 (2011).

\bibitem{weier2011quantum}
H.~Weier, H.~Krauss, M.~Rau, M.~F{\"u}rst, S.~Nauerth, and H.~Weinfurter, \enquote{Quantum eavesdropping without interception: an attack exploiting the dead time of single-photon detectors,} {\protect\JournalTitle{New Journal of Physics}} \textbf{13}, 073024 (2011).

\bibitem{huang2013quantum}
J.-Z. Huang, C.~Weedbrook, Z.-Q. Yin, S.~Wang, H.-W. Li, W.~Chen, G.-C. Guo, and Z.-F. Han, \enquote{Quantum hacking of a continuous-variable quantum-key-distribution system using a wavelength attack,} {\protect\JournalTitle{Physical Review A}} \textbf{87}, 062329 (2013).

\bibitem{yin2016measurement}
H.-L. Yin, T.-Y. Chen, Z.-W. Yu, H.~Liu, L.-X. You, Y.-H. Zhou, S.-J. Chen, Y.~Mao, M.-Q. Huang, W.-J. Zhang \emph{et~al.}, \enquote{Measurement-device-independent quantum key distribution over a 404 km optical fiber,} {\protect\JournalTitle{Physical review letters}} \textbf{117}, 190501 (2016).

\bibitem{tang2014field}
Y.-L. Tang, H.-L. Yin, S.-J. Chen, Y.~Liu, W.-J. Zhang, X.~Jiang, L.~Zhang, J.~Wang, L.-X. You, J.-Y. Guan \emph{et~al.}, \enquote{Field test of measurement-device-independent quantum key distribution,} {\protect\JournalTitle{IEEE Journal of Selected Topics in Quantum Electronics}} \textbf{21}, 116--122 (2014).

\bibitem{tang2016measurement}
Y.-L. Tang, H.-L. Yin, Q.~Zhao, H.~Liu, X.-X. Sun, M.-Q. Huang, W.-J. Zhang, S.-J. Chen, L.~Zhang, L.-X. You \emph{et~al.}, \enquote{Measurement-device-independent quantum key distribution over untrustful metropolitan network,} {\protect\JournalTitle{Physical Review X}} \textbf{6}, 011024 (2016).

\bibitem{rubenok2013real}
A.~Rubenok, J.~A. Slater, P.~Chan, I.~Lucio-Martinez, and W.~Tittel, \enquote{Real-world two-photon interference and proof-of-principle quantum key distribution immune to detector attacks,} {\protect\JournalTitle{Physical review letters}} \textbf{111}, 130501 (2013).

\bibitem{wang2022experimental}
C.~Wang, W.~Y. Kon, H.~J. Ng, and C.~C.-W. Lim, \enquote{Experimental symmetric private information retrieval with measurement-device-independent quantum network,} {\protect\JournalTitle{Light: Science \& Applications}} \textbf{11}, 268 (2022).

\bibitem{berrevoets2022deployed}
R.~C. Berrevoets, T.~Middelburg, R.~F. Vermeulen, L.~D. Chiesa, F.~Broggi, S.~Piciaccia, R.~Pluis, P.~Umesh, J.~F. Marques, W.~Tittel \emph{et~al.}, \enquote{Deployed measurement-device independent quantum key distribution and bell-state measurements coexisting with standard internet data and networking equipment,} {\protect\JournalTitle{Communications Physics}} \textbf{5}, 186 (2022).

\bibitem{pirandola2015high}
S.~Pirandola, C.~Ottaviani, G.~Spedalieri, C.~Weedbrook, S.~L. Braunstein, S.~Lloyd, T.~Gehring, C.~S. Jacobsen, and U.~L. Andersen, \enquote{High-rate measurement-device-independent quantum cryptography,} {\protect\JournalTitle{Nature Photonics}} \textbf{9}, 397--402 (2015).

\bibitem{li2014continuous}
Z.~Li, Y.-C. Zhang, F.~Xu, X.~Peng, and H.~Guo, \enquote{Continuous-variable measurement-device-independent quantum key distribution,} {\protect\JournalTitle{Physical Review A}} \textbf{89}, 052301 (2014).

\bibitem{ma2014gaussian}
X.-C. Ma, S.-H. Sun, M.-S. Jiang, M.~Gui, and L.-M. Liang, \enquote{Gaussian-modulated coherent-state measurement-device-independent quantum key distribution,} {\protect\JournalTitle{Physical Review A}} \textbf{89}, 042335 (2014).

\bibitem{papanastasiou2017finite}
P.~Papanastasiou, C.~Ottaviani, and S.~Pirandola, \enquote{Finite-size analysis of measurement-device-independent quantum cryptography with continuous variables,} {\protect\JournalTitle{Physical Review A}} \textbf{96}, 042332 (2017).

\bibitem{zhang2017finite}
X.~Zhang, Y.~Zhang, Y.~Zhao, X.~Wang, S.~Yu, and H.~Guo, \enquote{Finite-size analysis of continuous-variable measurement-device-independent quantum key distribution,} {\protect\JournalTitle{Physical Review A}} \textbf{96}, 042334 (2017).

\bibitem{lupo2018continuous}
C.~Lupo, C.~Ottaviani, P.~Papanastasiou, and S.~Pirandola, \enquote{Continuous-variable measurement-device-independent quantum key distribution: Composable security against coherent attacks,} {\protect\JournalTitle{Physical Review A}} \textbf{97}, 052327 (2018).

\bibitem{chen2018composable}
Z.~Chen, Y.~Zhang, G.~Wang, Z.~Li, and H.~Guo, \enquote{Composable security analysis of continuous-variable measurement-device-independent quantum key distribution with squeezed states for coherent attacks,} {\protect\JournalTitle{Physical Review A}} \textbf{98}, 012314 (2018).

\bibitem{ma2019long}
H.-X. Ma, P.~Huang, D.-Y. Bai, T.~Wang, S.-Y. Wang, W.-S. Bao, and G.-H. Zeng, \enquote{Long-distance continuous-variable measurement-device-independent quantum key distribution with discrete modulation,} {\protect\JournalTitle{Physical Review A}} \textbf{99}, 022322 (2019).

\bibitem{wilkinson2020long}
K.~N. Wilkinson, P.~Papanastasiou, C.~Ottaviani, T.~Gehring, and S.~Pirandola, \enquote{Long-distance continuous-variable measurement-device-independent quantum key distribution with postselection,} {\protect\JournalTitle{Physical Review Research}} \textbf{2}, 033424 (2020).

\bibitem{tian2022experimental}
Y.~Tian, P.~Wang, J.~Liu, S.~Du, W.~Liu, Z.~Lu, X.~Wang, and Y.~Li, \enquote{Experimental demonstration of continuous-variable measurement-device-independent quantum key distribution over optical fiber,} {\protect\JournalTitle{Optica}} \textbf{9}, 492--500 (2022).

\bibitem{hajomer2022high}
A.~A. Hajomer, H.~Q. Nguyen, U.~L. Andersen, and T.~Gehring, \enquote{High-rate continuous-variable measurement-device-independent quantum key distribution,} in \emph{Optical Fiber Communication Conference,}  (Optica Publishing Group, 2023), pp. M2I--2.

\bibitem{frohlich2015quantum}
B.~Fr{\"o}hlich, J.~F. Dynes, M.~Lucamarini, A.~W. Sharpe, S.~W.-B. Tam, Z.~Yuan, and A.~J. Shields, \enquote{Quantum secured gigabit optical access networks,} {\protect\JournalTitle{Scientific reports}} \textbf{5}, 18121 (2015).

\bibitem{wang2023experimental}
X.~Wang, Z.~Chen, Z.~Li, D.~Qi, S.~Yu, and H.~Guo, \enquote{Experimental upstream transmission of continuous variable quantum key distribution access network,} {\protect\JournalTitle{Optics Letters}} \textbf{48}, 3327--3330 (2023).

\bibitem{suleiman202240}
I.~Suleiman, J.~A.~H. Nielsen, X.~Guo, N.~Jain, J.~Neergaard-Nielsen, T.~Gehring, and U.~L. Andersen, \enquote{40 km fiber transmission of squeezed light measured with a real local oscillator,} {\protect\JournalTitle{Quantum Science and Technology}} \textbf{7}, 045003 (2022).

\bibitem{lupo2018parameter}
C.~Lupo, C.~Ottaviani, P.~Papanastasiou, and S.~Pirandola, \enquote{Parameter estimation with almost no public communication for continuous-variable quantum key distribution,} {\protect\JournalTitle{Physical review letters}} \textbf{120}, 220505 (2018).

\bibitem{jain2022practical}
N.~Jain, H.-M. Chin, H.~Mani, C.~Lupo, D.~S. Nikolic, A.~Kordts, S.~Pirandola, T.~B. Pedersen, M.~Kolb, B.~{\"O}mer \emph{et~al.}, \enquote{Practical continuous-variable quantum key distribution with composable security,} {\protect\JournalTitle{Nature communications}} \textbf{13}, 4740 (2022).

\bibitem{hajomer2022modulation}
A.~A. Hajomer, N.~Jain, H.~Mani, H.-M. Chin, U.~L. Andersen, and T.~Gehring, \enquote{Modulation leakage-free continuous-variable quantum key distribution,} {\protect\JournalTitle{npj Quantum Information}} \textbf{8}, 136 (2022).

\bibitem{gehring2021homodyne}
T.~Gehring, C.~Lupo, A.~Kordts, D.~Solar~Nikolic, N.~Jain, T.~Rydberg, T.~B. Pedersen, S.~Pirandola, and U.~L. Andersen, \enquote{Homodyne-based quantum random number generator at 2.9 gbps secure against quantum side-information,} {\protect\JournalTitle{Nature Communications}} \textbf{12}, 605 (2021).

\bibitem{chin2022digital}
H.-M. Chin, N.~Jain, U.~L. Andersen, D.~Zibar, and T.~Gehring, \enquote{Digital synchronization for continuous-variable quantum key distribution,} {\protect\JournalTitle{Quantum Science and Technology}} \textbf{7}, 045006 (2022).

\bibitem{minder2019experimental}
M.~Minder, M.~Pittaluga, G.~L. Roberts, M.~Lucamarini, J.~Dynes, Z.~Yuan, and A.~J. Shields, \enquote{Experimental quantum key distribution beyond the repeaterless secret key capacity,} {\protect\JournalTitle{Nature Photonics}} \textbf{13}, 334--338 (2019).

\bibitem{bordonalli1999high}
A.~Bordonalli, C.~Walton, and A.~J. Seeds, \enquote{High-performance phase locking of wide linewidth semiconductor lasers by combined use of optical injection locking and optical phase-lock loop,} {\protect\JournalTitle{Journal of Lightwave Technology}} \textbf{17}, 328 (1999).

\bibitem{kazovsky1989phase}
L.~G. Kazovsky, \enquote{Phase-and polarization-diversity coherent optical techniques,} {\protect\JournalTitle{Journal of Lightwave Technology}} \textbf{7}, 279--292 (1989).

\bibitem{zhang2020one}
Y.~Zhang, Y.~Huang, Z.~Chen, Z.~Li, S.~Yu, and H.~Guo, \enquote{One-time shot-noise unit calibration method for continuous-variable quantum key distribution,} {\protect\JournalTitle{Physical Review Applied}} \textbf{13}, 024058 (2020).

\bibitem{chin2021machine}
H.-M. Chin, N.~Jain, D.~Zibar, U.~L. Andersen, and T.~Gehring, \enquote{Machine learning aided carrier recovery in continuous-variable quantum key distribution,} {\protect\JournalTitle{npj Quantum Information}} \textbf{7}, 20 (2021).

\bibitem{marie2017self}
A.~Marie and R.~All{\'e}aume, \enquote{Self-coherent phase reference sharing for continuous-variable quantum key distribution,} {\protect\JournalTitle{Phys. Rev. A}} \textbf{95}, 012316 (2017).

\bibitem{walk2016experimental}
N.~Walk, S.~Hosseini, J.~Geng, O.~Thearle, J.~Y. Haw, S.~Armstrong, S.~M. Assad, J.~Janousek, T.~C. Ralph, T.~Symul \emph{et~al.}, \enquote{Experimental demonstration of gaussian protocols for one-sided device-independent quantum key distribution,} {\protect\JournalTitle{Optica}} \textbf{3}, 634--642 (2016).

\bibitem{gehring2015implementation}
T.~Gehring, V.~H{\"a}ndchen, J.~Duhme, F.~Furrer, T.~Franz, C.~Pacher, R.~F. Werner, and R.~Schnabel, \enquote{Implementation of continuous-variable quantum key distribution with composable and one-sided-device-independent security against coherent attacks,} {\protect\JournalTitle{Nature communications}} \textbf{6}, 8795 (2015).

\bibitem{bruynsteen2021integrated}
C.~Bruynsteen, M.~Vanhoecke, J.~Bauwelinck, and X.~Yin, \enquote{Integrated balanced homodyne photonic--electronic detector for beyond 20 ghz shot-noise-limited measurements,} {\protect\JournalTitle{Optica}} \textbf{8}, 1146--1152 (2021).

\bibitem{hajomer2024continuous}
A.~A. Hajomer, C.~Bruynsteen, I.~Derkach, N.~Jain, A.~Bomhals, S.~Bastiaens, U.~L. Andersen, X.~Yin, and T.~Gehring, \enquote{Continuous-variable quantum key distribution at 10 gbaud using an integrated photonic-electronic receiver,} {\protect\JournalTitle{Optica}} \textbf{11}, 1197--1204 (2024).

\end{thebibliography}

\bibliographyfullrefs{sample}


\ifthenelse{\equal{\journalref}{aop}}{%
\section*{Author Biographies}
\begingroup
\setlength\intextsep{0pt}
\begin{minipage}[t][6.3cm][t]{1.0\textwidth} 
  \begin{wrapfigure}{L}{0.25\textwidth}
    \includegraphics[width=0.25\textwidth]{john_smith.eps}
  \end{wrapfigure}
  \noindent
  {\bfseries John Smith} received his BSc (Mathematics) in 2000 from The University of Maryland. His research interests include lasers and optics.
\end{minipage}
\begin{minipage}{1.0\textwidth}
  \begin{wrapfigure}{L}{0.25\textwidth}
    \includegraphics[width=0.25\textwidth]{alice_smith.eps}
  \end{wrapfigure}
  \noindent
  {\bfseries Alice Smith} also received her BSc (Mathematics) in 2000 from The University of Maryland. Her research interests also include lasers and optics.
\end{minipage}
\endgroup
}{}

\end{document}